\documentclass[%
reprint,
superscriptaddress,
groupedaddress,
amsmath,amssymb,
aps,
pra,
]{revtex4-2}

\usepackage{graphicx}
\usepackage{dcolumn}
\usepackage{bm}
\usepackage{color}
\usepackage{qcircuit}
\usepackage{booktabs,eqparbox}
\usepackage{braket}
\usepackage{dsfont}
\usepackage{placeins}
\usepackage{comment}
\usepackage[utf8]{inputenc}
\usepackage{pgfplots}
\usepackage{hyperref}
\usepackage{cleveref}

\DeclareUnicodeCharacter{2212}{−}
\usepgfplotslibrary{groupplots,dateplot}
\usetikzlibrary{patterns,shapes.arrows}
\pgfplotsset{compat=newest}

\newcommand{\mh}[1]{{\color{red} #1}}

\begin{document}

\title{Tunable coupler to fully decouple and maximally localize superconducting qubits}

\author{Lukas Heunisch}
\email{lukas.heunisch@fau.de}
\affiliation{Physics Department, Friedrich-Alexander-Universität Erlangen Nürnberg, Germany}
\author{Christopher Eichler}
\affiliation{Physics Department, Friedrich-Alexander-Universität Erlangen Nürnberg, Germany}
\author{Michael J. Hartmann}
 \email{michael.j.hartmann@fau.de}
\affiliation{Physics Department, Friedrich-Alexander-Universität Erlangen Nürnberg, Germany}

\date{\today}

\begin{abstract}
Enhancing the capabilities of superconducting quantum hardware, requires higher gate fidelities and lower crosstalk, particularly in larger scale devices, in which qubits are coupled to multiple neighbors. Progress towards both of these objectives would highly benefit from the ability to fully control all interactions between pairs of qubits. Here we propose a new coupler model that allows to fully decouple dispersively detuned Transmon qubits from each other, i.e. $ZZ$-crosstalk is completely suppressed while maintaining a maximal localization of the qubits' computational basis states. We further reason that, for a dispersively detuned Transmon system, this can only be the case if the anharmonicity of the coupler is positive at the idling point. A simulation of a $40\,\mathrm{ns}$ CZ-gate for a lumped element model suggests that achievable process infidelity can be pushed below the limit imposed by state-of-the-art coherence times of Transmon qubits. On the other hand, idle gates between qubits are no longer limited by parasitic interactions. We show that our scheme can be applied to large integrated qubit grids, where it allows to fully isolate a pair of qubits, that undergoes a gate operation, from the rest of the chip while simultaneously pushing the fidelity of gates to the limit set by the coherence time of the individual qubits. 
\end{abstract}

\maketitle

\section{\label{sec:intro}Introduction}
Quantum computers are seen as a promising way to push the boundaries of computing power. In particular quantum chips based on superconducting qubits have proven to be a highly versatile platform and thus triggered significant efforts towards improving their performance \cite{AB, AC, AE, AL}. First experiments indicate that quantum computers can already outperform conventional supercomputers in specific benchmark tasks \cite{AA, AD}. 
Current devices are still limited by errors and noise \cite{AF},  such that currently achievable gate fidelities do not allow to solve practical problems of interest on a quantum device that cannot be computed on a classical computer. To break this barrier towards quantum advantage, two bottlenecks of current superconducting quantum hardware need to be improved upon: (i) The error rate of individual single- and two-qubit gates must be significantly reduced and (ii) interactions between idling qubits need to be further suppressed \cite{AW, AAB, AAC, AV, AX}. Indeed, such parasitic interactions can lead to uncontrollable quantum fluctuations, sometimes even to chaotic phases \cite{AG} for larger system sizes. 
\begin{figure}[h]
    \centering
    \includegraphics[width = \columnwidth]{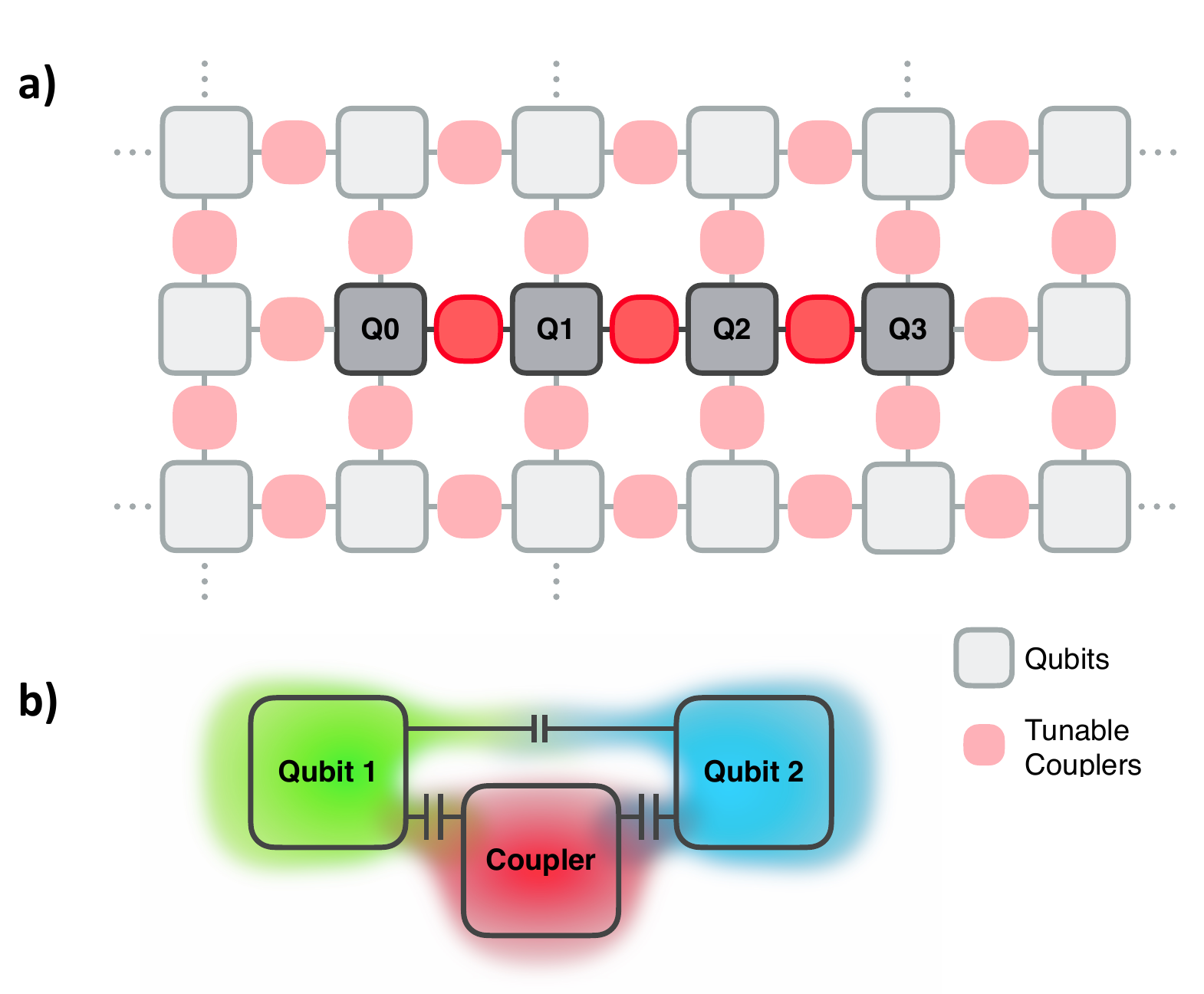}
    \caption{a) In large processors, qubits are often arranged in a square lattice and connected via tunable couplers for better control of the interactions between them. b) Sketch of such a tunable coupler circuit linking two qubits. Green, blue and red clouds indicate the wave functions of the two qubits (green and blue) and the coupler (red). The coupler scheme proposed in this work can achieve maximal localization of the computational basis and full suppression of this residual $ZZ$-crosstalk at the same time}
    \label{fig:standard_circuit}
\end{figure}
To control interactions between qubits, tunable coupler circuits have proven to be useful  (c.f. \autoref{fig:standard_circuit}) \cite{AR, AH, AK, AI, AJ, AM, AN}, since they allow to suppress the exchange of excitations between the circuit parts that form the qubits by destructive interference between the coupling mediated by the coupler and the direct capacitive coupling. An applied external magnetic flux allows to tune between regimes of strong and vanishing excitation exchange, where in the latter case the qubits are in idle. In many devices featuring Transmons, Transmon type circuits are also used as couplers \cite{AA, AD, AR, AH, AK, AI, AJ, AM, AN}. However, this has the disadvantage that longitudinal interactions ($ZZ$-crosstalk) remain present, even when the qubits are dispersively detuned and excitation exchange between them is strongly suppressed. 

Since $ZZ$-crosstalk causes phase errors \cite{AU, AS, AT}, there has been considerable effort to suppress such parasitic interactions \cite{BA, AO, AP, AQ, BC, BB, BE, BD, AY, AZ, AAD, AAE}. One approach is to couple two qubits with different sign anharmonicities \cite{AO, AP, AQ}, which however requires integrating two types of qubits into the same quantum processor in a staggered fashion. One could of course always suppress $ZZ$-crosstalk by detuning the qubits very strongly from each other at the idle point. To also be able to execute two-qubit gates, this strategy would however imply that  the qubit transition frequencies need to be tuned over large ranges, and thus significantly enhances the risk to couple to two-level system defects \cite{AAA}. 

While previous circuit architectures either achieve a suppression of excitation exchange at idle times by using a tunable coupler, or suppress $ZZ$-crosstalk, we propose a concept that can simultaneously optimize both effects.
We achieve this functionality by setting the idle point of the system at a point where the coupler has a different sign for its anharmonicity as compared to the Transmon qubits. We further show that by tuning a magnetic flux bias applied to the coupler, the device also generates strong interactions between the qubits, where numerical simulations for a lumped element model suggest that it allows to execute fast, high fidelity two-qubit gates. 

The coupling circuits we propose can lead to a significant improvement of the performance of multi-qubit superconducting circuit processors since they allow to isolate individual qubits that undergo a gate operation from the remaining qubits. The same of course holds when all qubits should be in idle. The benefits of being able to isolate qubits are also apparent from the fact that significantly higher gate fidelities can be achieved for individual qubit pairs than for qubits within a large integrated chip \cite{AA, AV, AW, AX}. Such switchable isolation of qubits is achieved by our design. 

The remainder of the paper is organized as follows. We first show how the coupler we propose can be employed to turn all interactions between a pair of qubits on and off. The quantities that we investigate for this purpose together with analytical approximations to them are presented in section \ref{sec:cross_analysis}, whereas the detailed proposed circuit for the coupler and numerically exact results for the decoupling are presented in section \ref{sec:model}. Then we present simulations for a fast, high-fidelity two-qubit CZ-gate in section \ref{sec:gatesimulation}, and finally show that the coupler can also achieve this functionality when it is part of a larger qubit lattice in section \ref{sec:integrated_system}, before closing with conclusions.

\section{\label{sec:cross_analysis} Crosstalk analysis}
In the following, we consider a two qubit system coupled by a tunable coupler. For our analysis, we describe the system by the Hamiltonian,
\begin{eqnarray}
\label{eqn:hamiltonian}
    H &=& \sum_{i=1,c,2} \omega_i b_i^\dagger b_i + \frac{U_i}{2} b_i^\dagger b_i^\dagger b_i b_i \\
    &+& g_{12} \bigl( b_1^\dagger b_2 + b_2^\dagger b_1\bigr) + \sum_{i=1, 2} g_{ic} \bigl( b_i^\dagger b_c + b_c^\dagger b_i \bigr) \ , \nonumber
\end{eqnarray}
where the qubits and couplers are described by anharmonic oscillators. The operators $b_i (b_i^\dag)$ for $i=1,2,c$ denote bosonic annihilation (creation) operators for the qubits ($i=1,2$) and the coupler ($i=c$). 

To precisely determine the crosstalk between two qubits, it is necessary to specify the experimentally relevant computational basis states. Denoting by $\ket{jkl}$ a local basis state of the chosen representation with $b_1^\dag b_1 \ket{jkl} = j \ket{jkl}$,  $b_c^\dag b_c \ket{jkl} = k \ket{jkl}$, and  $b_2^\dag b_2 \ket{jkl} = l \ket{jkl}$ for the ladder operators of \autoref{eqn:hamiltonian}, the corresponding experimentally relevant computational basis state is the eigenstate of the Hamiltonian $H$ in \autoref{eqn:hamiltonian}, for which the overlap with $\ket{jkl}$ is largest at the idle point, i.e. when the qubits are decoupled. We denote this state by $\ket{\widetilde{jkl}}$. 

We perform both, a full numerical and an analytical perturbative analysis of the model in \autoref{eqn:hamiltonian}, to investigate under which conditions we can reach a cancellation of residual $ZZ$-coupling while maintaining maximally localized states. From the analytical perturbative analysis, we derive a generic condition under which such a desirable operating point can be found. The numerically exact analysis provides us with precise values that are obtained without invoking approximations for the considered model and are plotted in the figures we present.

\subsection{\label{sec:XXcoupling} Localization of computational basis}
One expects a maximal decoupling of neighboring qubits when the states $\ket{\widetilde{100}}$ and $\ket{\widetilde{001}}$ are maximally localized in their respective circuits. To determine the parameter point at which this maximal localization occurs, one can compute overlaps between computational basis states $\ket{\widetilde{jkl}}$ and local basis states $\ket{jkl}$ and minimize the delocalization paramter 
\begin{equation}
\label{eqn:epsilon}
    \epsilon = \text{max} \left( \left| \langle \widetilde{100} | 001 \rangle \right|^2 \ , \ \left| \langle \widetilde{001} | 100 \rangle \right|^2 \right) \ .
\end{equation}
An approximate but analytical condition for the parameter point of minimal delocalization, min($\epsilon$), can be found via a standard time-independent perturbation expansion, see \autoref{eq:pert_comp_basis} in Appendix \ref{sec:statelocalization}.
Using a rotating wave approximation one finds that the perturbative approximation to $\epsilon$ becomes minimal for
\begin{equation}
    \label{eqn:SW_cond}
    g_{12} = - \frac{g_{1c}g_{2c}}{2}\left( \frac{1}{\Delta_{1c}} + \frac{1}{\Delta_{2c}}\right) \ ,
\end{equation}
where $\Delta_{ij} := \omega_i - \omega_j$ denotes the difference between the qubit frequencies. This condition (\ref{eqn:SW_cond}) can also be obtained via a Schrieffer-Wolff transformation \cite{DA} to a dressed mode picture, see Appendix \ref{sec:SWtrafo}. The Schrieffer-Wolff transformation yields an effective coupling $g_{\text{eff}}$ between the two qubits, that is mediated by the coupler. Requiring that $g_{\text{eff}} = 0$ one recovers condition (\ref{eqn:SW_cond}). Our approach of seeking maximal localization of the computational basis is thus in agreement with works that aim to suppress $g_{\text{eff}}$ \cite{AH}.

\subsection{\label{sec:ZZcoupling} Longitudinal coupling}
Since the computational basis is composed of eigenstates of the Hamiltonian $H$ at the idle point, the only residual coupling between the qubits at this point can be described as a $ZZ$-coupling. Such longitudinal couplings describe the dispersive shifts of the qubit energies resulting from the hybridization of the qubit wave functions due to the coupling capacitances. These can be quantified with the help of
\begin{equation}
    \label{eqn:zeta}
    \zeta = E_{101} - E_{100} - E_{001} + E_{000} \ ,
\end{equation}
where $E_{jkl}$ is the energy eigenvalue of the state $\ket{\widetilde{jkl}}$.  $\zeta$ can be calculated by truncating the Hamiltonian at excitation level number $N \gg \sqrt{2}q_{\text{ZPF}}/2e$ and diagonalizing it numerically, where $q_{\text{ZPF}}$ describe the charge zero point fluctuations of the Hamiltonian modes. 

To gain additional insights, an analytic but perturbative approach can be used to approximate $\zeta$ as defined in \autoref{eqn:zeta}. This analytic approach considers the qubit level spacing as an unperturbed Hamiltonian $H_0$, which is perturbed by the small interaction terms. Since $|g_{12}/\Delta_{12}| \ll |g_{1c}/\Delta_{1c}|, \ |g_{2c}/\Delta_{2c}|$, we keep up to second order terms in $g_{12}/\Delta_{12}$ and up to fourth order terms in $g_{1c}/\Delta_{1c}$ and $g_{2c}/\Delta_{2c}$ \cite{CE}, to approximate $\zeta$ as a sum of the relevant energy corrections of leading order, i.e. $\zeta \approx \sum_{i=1}^4 \zeta^{(i)}$ with $\zeta^{(i)} = E_{101}^{(i)} - E_{100}^{(i)} - E_{001}^{(i)} + E_{000}^{(i)}$. 
Hence the analytical approximate condition for vanishing  $ZZ$-crosstalk reads
\begin{equation}\label{eqn:zeta_perturbation}
      0 = \sum_{i=1}^4 \zeta^{(i)}  .
\end{equation}
We find that $\zeta^{(0)} = \zeta^{(1)} = 0$ and provide the explicit expressions for $\zeta^{(2)}, \zeta^{(3)}$ and $\zeta^{(4)}$, which are all of the same magnitude, in Appendix \ref{sec:ZZcouplingPert}.

\subsection{Maximal qubit-qubit decoupling}
We now show that both conditions, maximal localization of computational basis states as well as full suppression of $ZZ$-crosstalk can be satisfied simultaneously. We first investigate the approximate expressions to find an analytical condition for the parameters of this optimal idle point, before showing the  full decoupling between the qubits numerically for a specific coupler model. 

We find an analytical condition for maximal qubit-qubit decoupling by expressing the direct qubit-qubit coupling $g_{12}$ in Eq. (\ref{eqn:zeta_perturbation}) via the condition (\ref{eqn:SW_cond}). All remaining terms are proportional to $g_{1c}^2g_{2c}^2$ and can be made to mutually cancel each other for a suitable parameter choice. Solving the resulting equation for the coupler anharmonicity $U_c$ yields,
\begin{equation}
\label{eqn:Uc_lowcrosstalk}
    U_c = -\frac{U}{2} \left(1-\frac{U^2}{\Delta_{12}^2} - \frac{U}{2(\Delta_{1c} + \Delta_{2c})} \right)^{-1} ,
\end{equation}
where we assumed $U_1 = U_2 \equiv U$ for simplicity. A more general formula that accounts also for asymmetries of the qubit anharmonicities can be found in Appendix \ref{sec:ZZsuppression1}. As one can see, $ZZ$-crosstalk can be completely suppressed within the accuracy of this perturbative treatment for qubits in the straddling regime ($U > \Delta_{12}$) if the coupler anharmonicity has the same sign as the anharmonicity of the qubits, c.f. \cite{AI}. However, if the qubit detuning becomes sufficiently large, $ZZ$-crosstalk can only be suppressed if the coupler has the opposite sign anharmonicity compared with the qubits.

We note that, similar expressions for cross-talk suppression have been obtained via second order Schrieffer-Wolf transformations in \cite{AAE}. 
Due to the necessary circuitry for connecting qubits, the delocalization of the computational basis is however a stronger effect than residual crosstalk and appears in higher order of perturbation theory than the latter. Whereas residual crosstalk directly leads to gate infidelities, including for idle gates, delocalized computational basis modes can largely be accounted for in calibration. Our approach achieves full suppression of the deterimental crosstalk and minimization of the more tolerable delocalization at the same time.

Since $\Delta_{12} = \Delta_{1c} - \Delta_{2c}$ there are two independent parameters $\Delta_{1c}$ and $\Delta_{2c}$,  which  can be adjusted such that \autoref{eqn:Uc_lowcrosstalk} and \autoref{eqn:SW_cond} are simultaneously satisfied. However, in reality the choice of qubit transition frequencies may be restricted, such that the coupler frequency remains the only free adjustable parameter. To nonetheless be able to satisfy \autoref{eqn:Uc_lowcrosstalk} and \autoref{eqn:SW_cond} the anharmonicity of the coupler needs to be implemented appropriately, which could be challenging due to the finite precision of the fabrication process. In \autoref{sec:ZZsuppression2} we therefore estimated the effect of imprecisions in the coupler fabrication and find that this only leads to very small residual $ZZ$-coupling.

The results of the perturbative calculations at the order employed here, i.e. Equations \ref{eqn:SW_cond}, \ref{eqn:zeta_perturbation} and \ref{eqn:Uc_lowcrosstalk}, remain valid if the considered qubits couple to further spectator qubits, as they would if they are part of a larger qubit grid. We further investigate this property via exact numerics in \autoref{sec:integrated_system}. 

In the following section, we will build on these findings to propose a design for a coupler for two Transmons that satisfies \autoref{eqn:Uc_lowcrosstalk}.

\section{\label{sec:model}C-Shunt Flux Coupler}
\begin{figure}
    \centering
    \includegraphics[width = \columnwidth]{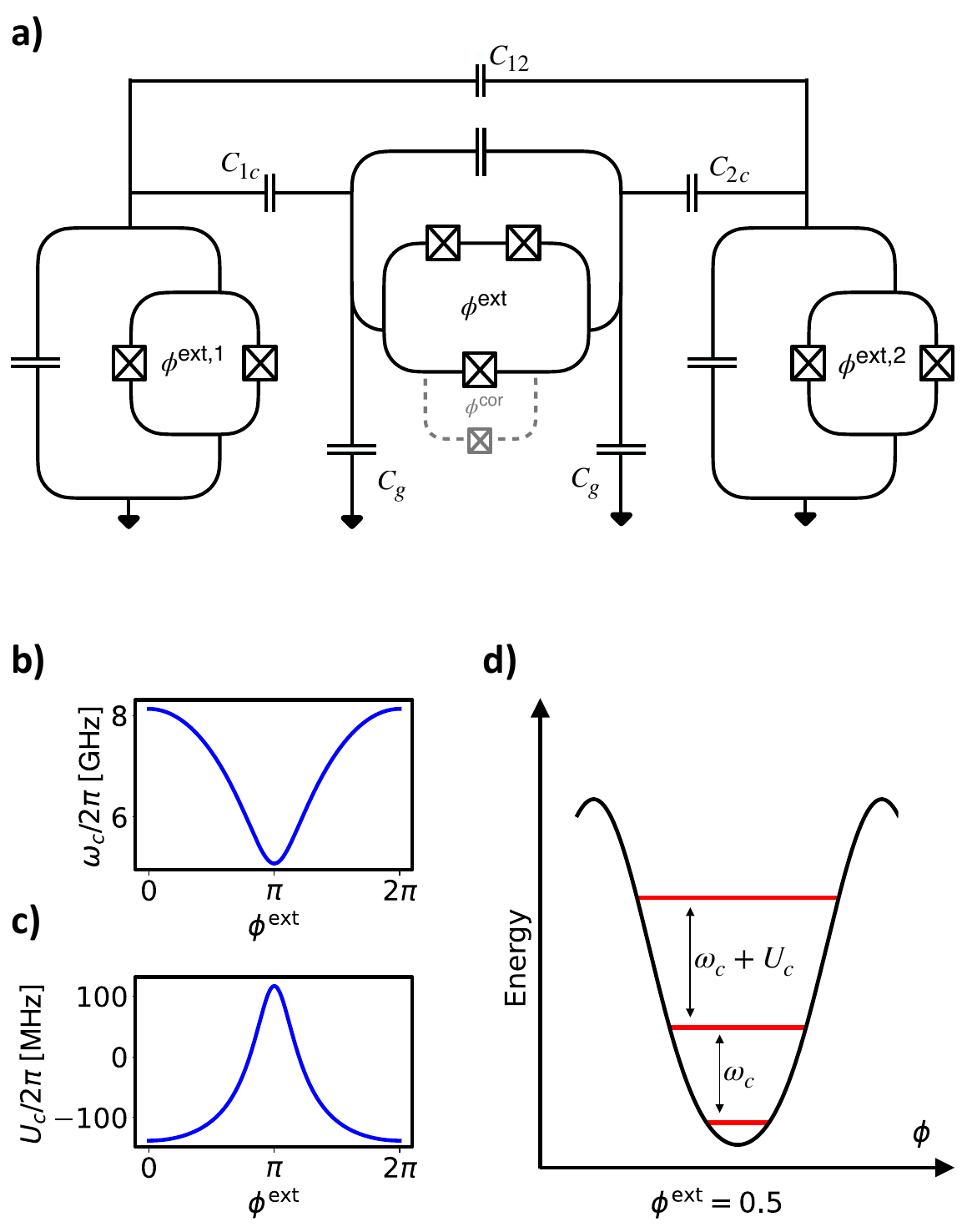}
    \caption{a) Circuit diagram of the investigated system. Two tunable Transmons are coupled via a C-shunt flux coupler. The positive anharmonicity of the C-shunt flux coupler can lead to simultaneous cancellation of transverse and longitudinal couplings if the parameters are chosen appropriately. To enable compensation of fabrication inaccuracies in calibration, the smaller junction in the bottom branch of the coupling circuit can be fabricated as a parallel pair of junctions (dashed) to make it tunable via an additional magnetic flux bias $\phi^{\text{cor}}$. b)-d): Energy spectrum of the coupler. Tuning the magnetic flux $\phi^{\text{ext}}$ provides not only a change in the coupler frequency $\omega_c$ but also a change in the anharmonicity $U_c$. For the simulation the parameters in Appendix \ref{sec:parameters_detailed} were used.}
    \label{fig:spectrum}
\end{figure}
We consider a setup, in which two tunable Transmons are coupled via a tunable coupler formed by a C-shunt flux qubit \cite{DD, DE, DC}. This circuit design is intended to achieve that two Transmons outside of the straddling regime can be completely decoupled due to the positive anharmonicity of the coupler. The diagram for this circuit, which we describe on the level of lumped elements \cite{DB}, is shown in \autoref{fig:spectrum}. The coupling element is grounded via the capacitances $C_g$. The qubits are directly coupled via capacitance $C_{12}$ and connected to the coupler via the capacitances $C_{1c}$ and $C_{2c}$. The capacitance $C_{12}$ was chosen to be an order of magnitude smaller than $C_{ic}$, which carries over to the coupling constants $g_{12} \ll g_{1c}, \ g_{2c}$, which describe direct coupling between the qubits and coupler mediated interaction, respectively. 

The effective interaction between the qubits is determined by the coupler, whose potential can be described using the fluxoid quantization condition by
\begin{equation}
\label{eqn:couplerPotential}
    V = - 2E_{J,c} \cos \left( \frac{\phi_c}{\sqrt{2}}\right) - \alpha E_{J,c} \cos \left( \sqrt{2} \phi_c + \phi^{\text{ext}} \right),
\end{equation}
where the parameter $\alpha$ denotes the ratio of the Josephson energies $E_{J,c}$ of the junctions in the upper branch and the junction in the lower branch of the coupler. The phase variable $\phi_c \equiv \phi_- = 1/\sqrt{2} \cdot (\phi_{c1} - \phi_{c2})$ is the difference of the active nodes $\phi_{1c}$ and $\phi_{2c}$. With the choice $1/8 < \alpha < 1/2$ the potential lies in its single-well regime, so that it can be approximated for small zero point fluctuations of $\varphi_c = \phi_c - \phi_{\text{min}}$ with the help of a Taylor expansion in $\varphi_c$ around the potential minimum $\phi_{\text{min}}$. The full derivation of the resulting Hamiltonian and explicit expressions of its parameters are provided in Appendix \ref{sec:circuitderivation}. After a rotating wave approximation, which we did not apply in  our numerical analysis, this circuit can be described by a Hamiltonian as in \autoref{eqn:hamiltonian}. 

By varying the control parameter $\phi^{\text{ext}} = 2\pi \Phi^{\text{ext}}/\Phi_0$, the potential and thus the level structure of the coupler can be manipulated as shown in \autoref{fig:spectrum}.  Note that positive anharmonicities always occur along with low frequencies in the coupler, motivating the choice of a floating C-shunt flux qubit as the coupling element, see \autoref{fig:spectrum}. Since the coupler mode can be identified with the physical $\phi_-$-mode, the constants $g_{1c}$ and $g_{2c}$ have different signs. Thus, the idle point of the coupler is below the qubit frequencies, which makes it possible to exploit the positive anharmonicity at a small coupler frequency, while one can also simply switch on effective interactions to execute gates by increasing the coupler frequency. 

The dynamics of the proposed circuit architecture is generated by its Hamiltonian as derived in Appendix \ref{sec:circuitderivation}, c.f. \autoref{eqn:hamiltonian}.

\subsection*{\label{sec:parameters} Parameter choice for the simulation}
To achieve perfect cancellation of $ZZ$-crosstalk at the point of maximum state localization, suitable circuit parameters need to be determined. We set the qubit transition frequencies to $\omega_1/2\pi = 6.6\,\mathrm{GHz}$ and $\omega_2/2\pi = 6.1\,\mathrm{GHz}$. The coupler level should be below the qubit levels at the idling point, where its positive anharmonicity can cause the cancellation of $ZZ$-crosstalk. 

\begin{figure}[!h]
    \centering
    \includegraphics[width = \columnwidth]{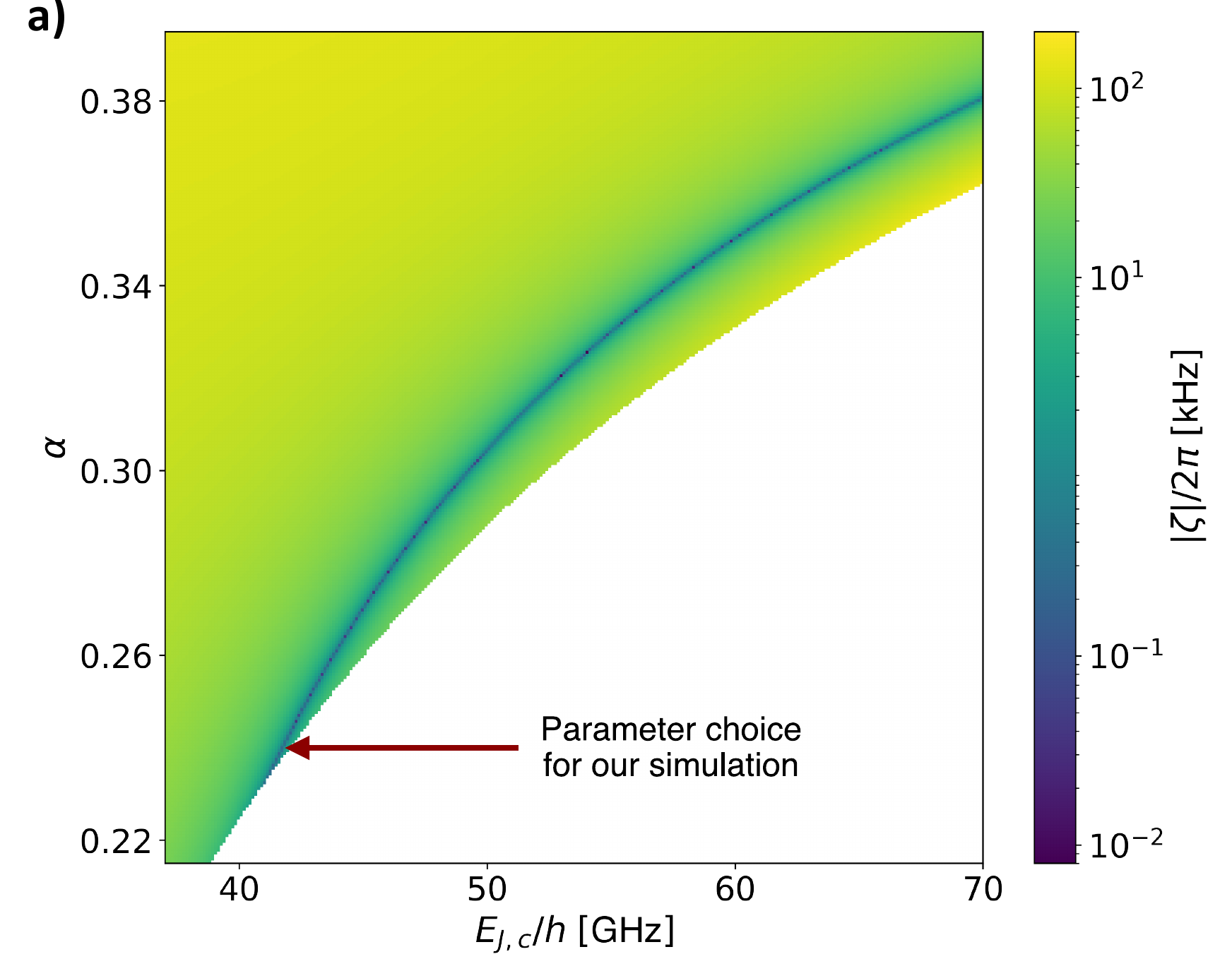}
    \includegraphics[width = \columnwidth]{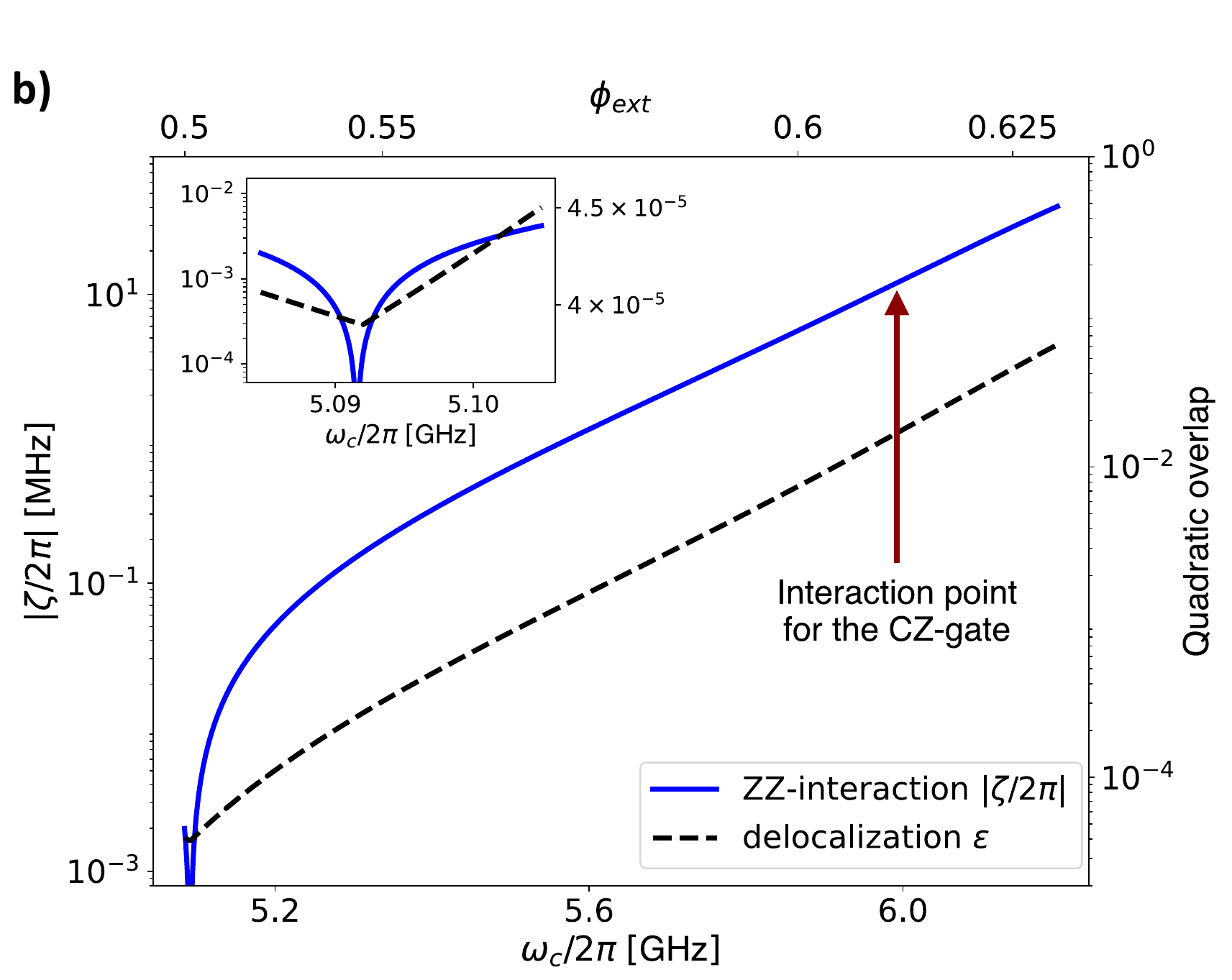}
    \caption{\label{fig:interactions} a) Residual $ZZ$-coupling $\zeta$ in kHz at the point of maximally localized states for various parameter constellations of the presented tunable coupler scheme. The dark blue line shows the area where $ZZ$-coupling can be completely suppressed without accepting a higher delocalization of eigenstates at the idle point. b) Effective longitudinal coupling $\zeta$ and qubit wave function delocalization $\epsilon$ numerically computed from Equations (\ref{eqn:epsilon}) and (\ref{eqn:zeta}), as a function of coupler frequency. For the parameter choice $E_J/h = 41.2\,\mathrm{GHz}$ and $\alpha = 0.2347$ used here, it is possible to switch off longitudinal interactions at the point of maximal qubit wave function localization tuning the coupler frequency to $\omega_c/2\pi \approx 5.092\,\mathrm{GHz}$. Since $\phi_{\text{ext}} = 0.5$ belongs to point of the lowest possible coupler frequency, the idling point can be seen at the far right edge of the plot.}
\end{figure}

For the C-shunt flux coupler, we assume identical Josephson junctions in the upper branch, while the junction in the lower branch is smaller by a factor $\alpha$. Since we want to fix both frequency and anharmonicity of the coupler at the idle point of the system, $\alpha$ and $E_{J,c}$ must be chosen suitably. This either needs to be achieved in the fabrication process or the small junction can be fabricated as a dc-SQUID to make it tunable via an additional magnetic flux bias, see \autoref{fig:spectrum}. We determined possible parameter choices numerically, see \autoref{fig:interactions}. As can be seen in the top panel, there is a sizable parameter range for the junction parameters of the Josephson junctions in the coupler, where the qubit states are maximally localized and $ZZ$-crosstalk can be fully suppressed, see blue line in the plot. For parameter sets on this line, it is possible to have a crosstalk-free idle point in the system by adapting the external flux.
By varying the external flux away from the idle point, interactions between the qubits can be switched on. For the simulation in the following sections, we have chosen a small Josephson energy of $E_{J}/h = 41.2\,\mathrm{GHz}$, because for such a parameter choice the slope of $\omega_c(\phi_{\text{ext}})$ is small and allows to increase the robustness of the coupler against flux noise at the $Z$-drive line. \autoref{fig:interactions} b) shows that $ZZ$-crosstalk and qubit state delocalization can be minimized simultaneously for this parameter choice. For a coupler frequency of $\omega_c/2\pi \approx 5.092\,$GHz the qubits can be completely decoupled. Increasing the coupler frequency leads to effective couplings, which can be used to perform high-fidelity two-qubit gates as we will show in \autoref{sec:gatesimulation}. A detailed compilation of all cicuit parameters can be found in Appendix \ref{sec:parameters_detailed}.

\subsection*{\label{sec:ZZsuppression2} Robustness against deviations in circuit parameters}
In this section we investigate the robustness of this regime of full decoupling against imperfections in the fabrication process of the circuits. To estimate the effects parameter fluctuationson residual cross-talk and delocalization, we numerically explored these for the \mh{above presented} example. We are interested in the requirements for the coupling circuit and consider relative errors in its charging and Josephson energies. I.e., denoting their optimal values by $E_{C, \text{opt}}$ and $E_{J_C, \text{opt}}$, we consider $\delta E_C$ and $\delta E_{J} = \text{sgn}(\delta E_{J_C}) \sqrt{\delta E_{J_C}^2 + \delta (\alpha E_{J_C})^2}$, where $\delta E_\nu = 1 - E_\nu/E_{\nu, \text{opt}}$ for $\nu = C, J_C$. Here, the quantity $\delta E_{J_C}$ was chosen to estimate the effects of fabrication errors in the junction in the lower branch as well as the two junctions of the upper branch via a single quantity. The sign of the errors for the smaller junction in the lower branch of the coupler $\delta (\alpha E_{J_C})$ and the two junctions in the upper branch $\delta E_{J_C}$ was thereby chosen differently in order to obtain an upper limit for the extent of the fabrication errors. The optimal values are thereby given by the values found in \autoref{sec:parameters}.

\begin{figure*}[!htp]
    \centering
    \includegraphics[width = 12.9cm]{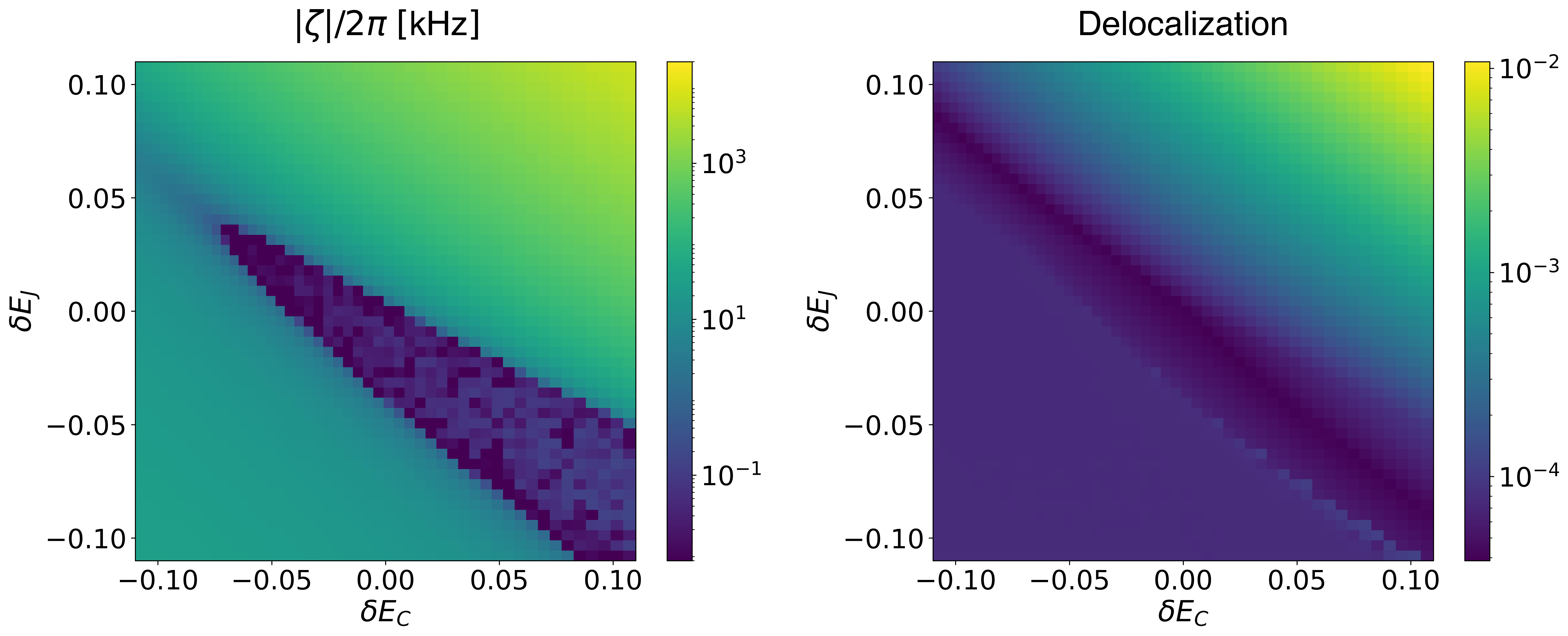}
    \caption{Numerical simulation of the $ZZ$-crosstalk (left) and state delocalization (right) in the presence of fabrication inaccuracies  $\delta E_C$ and $\delta E_{J}$. The magnetic flux $\Phi_c^{\text{ext}}$ was initialized such that the $ZZ$-crosstalk is minimized. A broad regime can be seen in which $ZZ$-crosstalk can be completely suppressed while accepting just a small state delocalization.}
    \label{fig:ZZ_deviation2}
\end{figure*}

We explore the effect of these fabrication inaccuracies on residual cross-talk and delocalization in \autoref{fig:ZZ_deviation2}, where the magnetic flux is adjusted such that $ZZ$-crosstalk is minimal. As \autoref{fig:ZZ_deviation2} shows, there is a broad regime, in which $ZZ$-crosstalk can be completely suppressed while  delocalization remains at a comparable level as in the ideal case, i.e. quadratic overlap $\sim 10^{-4}$ (c.f. \autoref{fig:interactions}). This illustrates the robustness of the proposed scheme against fabrication errors.

Additionally, one could compensate fabrication inaccuracies in calibration, by replacing the smaller junction in the coupling circuit by a flux-tunable superconducting quantum interference device (SQUID) loop. Controlling the magnetic flux allows for tuning  the effective Josephson energy of the SQUID. The susceptibility to flux noise of this element can furthermore be reduced by using two junctions with different Josephson energies, since the effective Josephson energy of the element then only weakly depends on the external flux.

\section{\label{sec:gatesimulation} Simulation of a CZ-gate}
We simulated a CZ-gate \cite{CF} on the circuit model presented in \autoref{sec:model}, using the QuTip tool \cite{CA, CB} for python and the $\mathds{C}^3$ tool \cite{CC} to optimize the pulse shapes.  The CZ-gate can be implemented by tuning the coupler to higher frequencies. This switches on a $ZZ$-interaction between the qubits as shown in \autoref{fig:interactions} b), which gives rise to a phase $\pi$ for the $\ket{\widetilde{101}}$ state. Holding the coupler frequency for a sufficiently large time at the point of strong $ZZ$-interactions, the target unitary $U_{\text{CZ}} = \text{diag}(1, 1, 1, -1)$ can be achieved up to single qubit phases, which can be compensated virtually. We chose this gate scheme because it allows to keep the qubit frequencies constant up to dispersive shifts due to the $ZZ$-interaction. This approach therefore allows neighboring spectator qubits, coupled by further C-shunt flux couplers, to remain fully decoupled during the gate operation. This mechanism thus avoids possible deterioration of gate fidelities caused by spectator qubits. 

For our simulation we use Gaussian flattop pulses for frequency tuning in order to suppress leakage state population at the end of the gate. While the  rounded pulse shapes cause the population of the leakage states to decrease at the end of the gate, the leakage population during the gate is determined by the detuning of the neighboring energy levels. In total, we simulate a CZ-gate with a length of $40\,\mathrm{ns}$. Its process infidelity is defined by 
\begin{equation}
\label{eqn:infidelity}
    \varepsilon = 1 - f = 1 - \frac{1}{4} \, \left| \text{Tr} \ U_{\text{CZ}}^\dagger U_{\text{simulation}} \right|
\end{equation}
and can be suppressed in this simulation for the optimal control pulses to $2.2 \cdot 10^{-4}$. To quantify leakage we have computed the ocupation probabilities of leakage states after the gate and found that, for computational basis states as inputs, this never exceeds $6 \cdot 10^{-4}$. Thus, the systematic error of the gate under the approximations that were made in this simulation is lower than the error due to decoherence. The latter can be estimated for state of the art Transmons with a coherence time of about $\tau \approx 50\,\mu\mathrm{s}$ \cite{AI} to $\varepsilon_{\text{decoh.}} = 1 -\exp (-t_{\text{gate}}/\tau)\approx 8 \cdot 10^{-4}$. Similar coherece times have been reported for C-shunted flux qubit circuits \cite{DC}, similar to the coupler circuit we consider.

We note that our coupler model would also be suited for other strategies to implement CZ-gates, e.g. by bringing the states $\ket{\widetilde{101}}$ and $\ket{\widetilde{200}}$ into resonance, where the phase $\pi$ is generated by a full Rabi oscillation between the two states. We analyze this method in Appendix \ref{sec:CZ_fast_ad}.

\section{\label{sec:integrated_system} Larger Clusters of qubits}
To show that our strategy also allows to fully decouple selected qubits in larger grids, we study a chain of four qubits as shown in \autoref{fig:standard_circuit}. Our aim here is to show that design parameters, such as the Jospehson energies of the junctions, which enable an optimal idle point in an isolated qubit dimer, c.f. \autoref{fig:interactions}, can be used in a large qubit grid to fully decouple selected qubit pairs via a suitable choice of the transition frequency or flux bias for the coupler between them. 

We thus investigate localization of computational basis wave functions and $ZZ$-crosstalk for the four qubit chain in \autoref{fig:standard_circuit}, where the coupler parameters (Josephson energies and capacitances) are chosen such that two adjacent qubits fully decouple if isolated from the remaining qubits. While for Q1 and Q2 we took the parameters found in \autoref{sec:parameters}, the frequencies $\omega_0$ and $\omega_3$ were fixed to 5.9 and 6.5\,GHz, respectively. Since Q1 and Q2 have here two neighboring qubits each, their shunt capacitances were adjusted such that their charging energies $E_C$ remained the same as for the isolated dimer despite the increased number of coupling capacitances connected to them. This ensures that the results for the 4-qubit chain are comparable with those of the isolated dimers.
\begin{figure*}
    \centering
    \includegraphics[width = \textwidth]{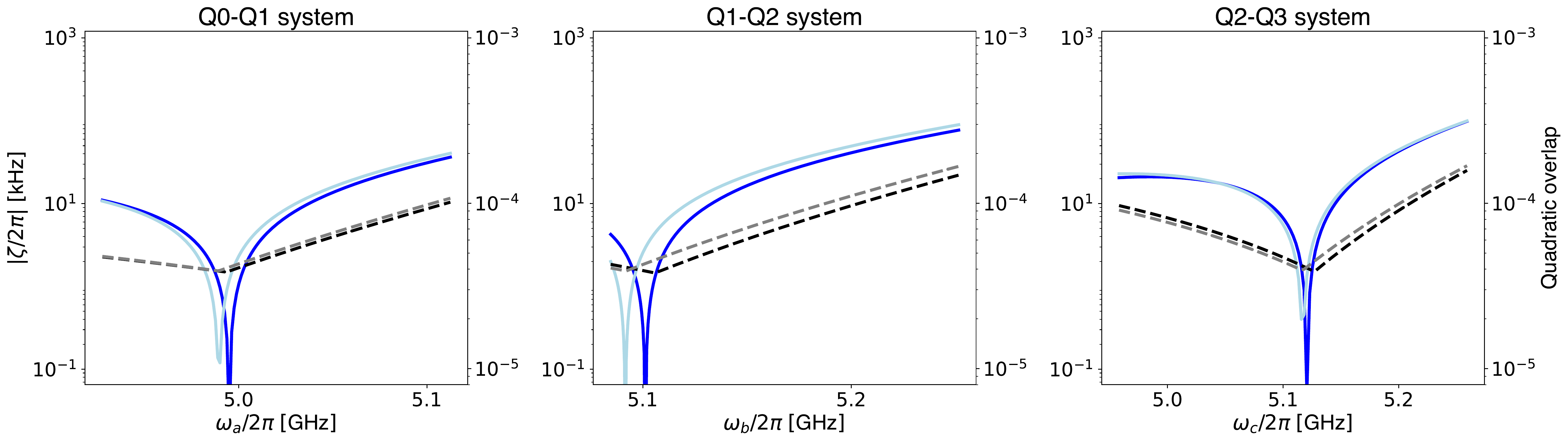}
    \caption{Qubit decoupling in a four qubit chain. The dark blue lines represent $ZZ$-crosstalk between Q0 and Q1, Q1 and Q2 and Q2 and Q3 respectively, the black dashed line represents qubit state delocalization. For comparison the light blue line shows $ZZ$-crosstalk and the gray dashed line qubit state delocalization for the respective qubit pairs when investigating them in isolation.}
    \label{fig:4Qubitsystem}
\end{figure*}

As mentioned already in \autoref{sec:cross_analysis}, the perturbative estimates indicate that full decoupling of neighboring qubits occurs in the grid for the same parameters as in the isolated dimer. In a numerically exact analysis we find that slight modifications of the flux biases at the couplers are needed to precisely hit the optimal idle points. \autoref{fig:4Qubitsystem} shows $ZZ$-crosstalk (dark blue) and state localization (black dashed) in the presence of spectator qubits for the qubit pairs Q0 - Q1, Q1 - Q2, and Q2 - Q3. For comparison, the same quantities have been plotted for the case where the qubits are considered in isolation from their spectators (light blue: $ZZ$-crosstalk, gray dashed: state localization). As can be seen, the points of maximal basis localization and full $ZZ$-crosstalk suppression coincide very accurately for both cases, while the presence of spectator qubits only cause a slight shift of the optimal idle points to slightly different coupler frequencies. 

Our investigation thus shows that interactions beyond neighboring qubits, which strongly decay with the distance between qubits but are nonetheless present, as well as interactions between couplers have a small effect on the parameters for optimal idle points, where neighboring qubits fully decouple. The slight shifts of the optimal idle points, that are caused by spectator qubits, can be compensated for by a minor adjustment of the coupler frequency. We conclude from these results that our approach to fully decouple neighboring qubits can also be applied to large qubit grids and should thus enable improved perfomance of large superconducting quantum processors.

\section{\label{sec:conclusions}Conclusions}
In this work, we have determined criteria, which tunable couplers should fulfill to enable a full decoupling of neighboring qubits in superconducting processors.  Whereas the computational basis states can be maximally localized in the respective qubit circuits by a suitable choice of the coupler frequency, we find that the remaining $ZZ$-interactions can be simultaneously suppressed if the sign of the coupler anharmonicity is chosen oppositely to that of the qubits. We showed that this concept applies to both, isolated qubit dimers as well as qubit pairs within larger grids. 

For Transmon type qubits, this arrangement can be achieved with C-shunt flux coupling circuits. For this architecture, we showed via simulations of a lumped element model that CZ-gates which are only limited by decoherence should be achievable. Alternatively our strategy can also be applied to C-shunt flux qubits that are coupled by Transmon type tunable couplers. 

Our work suggests that using coupling circuits with suitable, opposite sign nonlinearities as compared to the qubits should enable significant improvements in the performance of quantum processors.

\begin{acknowledgments}
The authors thank Verena Feulner for helpful discussions in early stages of this project and Federico Roy for help in using the $\mathds{C}^3$ tool. This work has received funding from the German Federal Ministry of Education and Research via the funding program quantum technologies - from basic research to the market under contract numbers 13N15684 "GeQCoS" and 13N16182 "MUNIQC-SC". It is also part of the Munich Quantum Valley, which is supported by the Bavarian state government with funds from the Hightech Agenda Bayern Plus.
\end{acknowledgments}

\appendix
\section{\label{sec:circuitderivation} Circuit Hamiltonian derivation}
\begin{figure}[!h]
    \centering
    \includegraphics[width = \columnwidth]{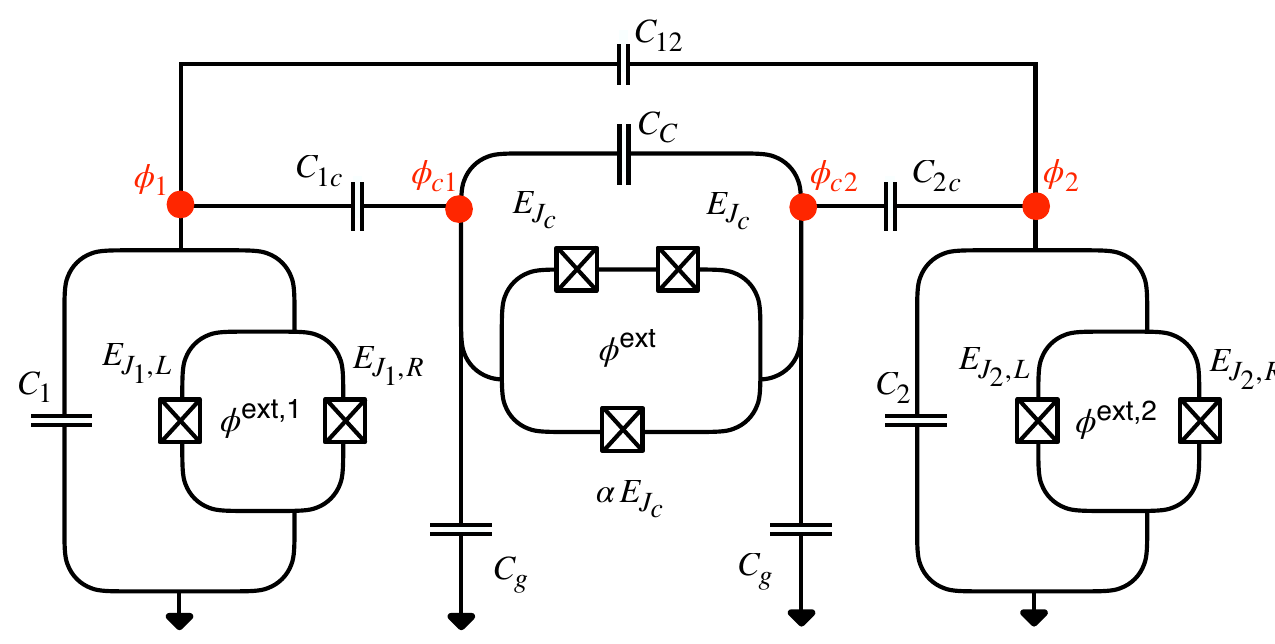}
    \caption{Circuit diagram of two tunable Transmons coupled via a C-Shunt flux coupler.}
    \label{fig:circuit2}
\end{figure}

We describe the circuit shown in \autoref{fig:circuit2} on the level of lumped elements using circuit quantum electrodynamics. The dynamics of this circuit is determined by the active nodes which are marked red. The Lagrangian is given by 
\begin{equation}
\label{eqn:lagrangian}
    \mathcal{L} = T - V \ ,
\end{equation}
where the kinetic energy $T$ is given by 
\begin{equation}
    T = \frac{1}{2} \dot{\mathbf{\Phi}}^T \mathbf{C} \dot{\mathbf{\Phi}} \ .
\end{equation}
The vector $\mathbf{\Phi}^T = (\phi_1, \phi_{1c}, \phi_{2c}, \phi_2)$ contains all node fluxes and the capacitance matrix $\mathbf{C}$ is given by:
\begin{widetext}
\begin{equation}
    \mathbf{C} = \begin{pmatrix}
        C_1 + C_{1c} + C_{12} & -C_{1c} & 0 & -C_{12} \\
        -C_{1c} & C_{1c} + C_g + C_C & -C_C & 0 \\
        0 & -C_C & C_{2c} + C_C + C_g & -C_{2c} \\
        -C_{12} & 0 & -C_{2c} & C_2 + C_{2c} + C_{12} 
    \end{pmatrix}
\end{equation}
\end{widetext}

We apply a transformation to our basis of the form 
\begin{equation}
    \mathbf{S} = \begin{pmatrix}
        1 & 0 & 0 & 0 \\
        0 & 1/\sqrt{2} & 1/\sqrt{2} & 0 \\
        0 & 1/\sqrt{2} & -1/\sqrt{2} & 0 \\
        0 & 0 & 0 & 1
    \end{pmatrix} \ ,
\end{equation}
which ensures on the one hand, that the capacitance matrix is diagonal dominated for $C_1, C_2, C_C, C_g \gg C_{1c}, C_{2c} \gg C_{12}$, which allows a proper definition of the qubit energy spacing and guarantees on the other hand, that the condition for the canonically conjugate variables $\phi_i$ and $q_i = \partial \mathcal{L}/\partial \dot{\phi}_i$ 
\begin{equation}
    \left[ \phi_i, \, q_i\right] = i
\end{equation}
is still satisfied. The potential $V$ in \autoref{eqn:lagrangian} is given by the sum of the qubit and coupler potentials. While the Transmon potentials are simply given by 
\begin{eqnarray}
    V_i &=& E_{J_i}^{\text{eff}}(\phi_i^{\text{ext}}) \cos \phi_i \nonumber \\
    &\approx & E_{J_i}^{\text{eff}}(\phi_i^{\text{ext}}) \left( -1 + \frac{\phi_i^2}{2} - \frac{\phi_i^4}{24} \right) \ ,
\end{eqnarray}
where the effective Josephson energy $E_{J_i}^{\text{eff}}$ is given by
\begin{eqnarray}
    E_{J_i}^{\text{eff}}(\phi_i^{\text{ext}}) & &= (E_{J_i,L} + E_{J_i,R}) \ \times \nonumber \\
    & &\sqrt{\cos ^2\left( \frac{\phi_i^{\text{ext}}}{2}\right) + d_i^2 \sin ^2\left( \frac{\phi_i^{\text{ext}}}{2} \right)} \ ,
\end{eqnarray}
where $d_i = \frac{E_{J_i,L} - E_{J_i,R}}{E_{J_i,L} + E_{J_i,R}}$ denotes the junction asymmetry of the qubit and $\phi_i^{\text{ext}} = 2\pi \Phi_i^{\text{ext}}/\Phi_0$ is the external parameter, which can be controlled by a magnetic flux. 

The coupler potential reads (see \autoref{eqn:couplerPotential})
\begin{align}
    V = &- 2E_{J_C} \cos \left( \frac{\phi_c}{\sqrt{2}}\right) \notag \\
    &- \alpha E_{J_C} \cos \left( \sqrt{2} \phi_c + \phi^{\text{ext}} \right) \ ,
\end{align}
where $\phi_c \equiv \phi_-$ is the coupler mode coming from the basis change $\tilde{\mathbf{\Phi}} = \mathbf{S}\mathbf{\Phi}$. The selfcapacitances of the junctions were absorbed in $C_1$, $C_2$, and $C_C$, respectively. For small zero point fluctuations of $\phi_c$ we can expand the potential around the minimum $\phi_{\text{min}}$ 
\begin{eqnarray}
V &\approx & V(\phi_{\text{min}}) \nonumber \\ 
&+& \frac{E_{J_C}}{2}\Biggl( \cos \left( \frac{\phi_{\text{min}}}{\sqrt{2}} \right) + 2\alpha \cos \left( \sqrt{2}\phi_{\text{min}} + \phi^{\text{ext}} \right) \Biggr) \varphi_c^2 \nonumber \\
&-& \frac{E_{J_C}}{6}\Biggl( \frac{1}{\sqrt{2}} \sin \left( \frac{\phi_{\text{min}}}{\sqrt{2}} \right) \nonumber \\ 
& & \ \phantom{\frac{E_{J_C}}{6}\Biggl(} + 2\sqrt{2}\alpha \sin \left( \sqrt{2}\phi_{\text{min}} + \phi^{\text{ext}} \right) \Biggr) \varphi_c^3 \nonumber \\
&-& \frac{E_{J_C}}{24}\Biggl( \frac{1}{2} \cos \left( \frac{\phi_{\text{min}}}{\sqrt{2}} \right) \nonumber \\ 
& & \ \phantom{\frac{E_{J_C}}{6}\Biggl(} + 4\alpha \cos \left( \sqrt{2}\phi_{\text{min}} + \phi_{\text{ext}} \right) \Biggr) \varphi_c^4 \ ,
\end{eqnarray}
where we defined $\varphi_c \equiv \phi_c - \phi_{\text{min}}$. The prefactors of the powers of $\varphi_c$ can only be found numerically, since the calculation of them requires solving the transcendental equation $\sin (\phi_c/\sqrt{2}) + \alpha \sin (\sqrt{2}\phi_c + \phi^{\text{ext}}) = 0$ which locates the minimum of the potential as a function of $\phi^{\text{ext}}$. 

Finding the Hamiltonian of the system requires finding the inverse capacitance matrix $\mathbf{C}^{-1}$, since the canonical conjugate momenta $q_i = \partial \mathcal{L}/\partial \phi_i$ are given by 
\begin{equation}
    \mathbf{q} = \mathbf{C} \dot{\mathbf{\Phi}} \ .
\end{equation}
Performing the Legendre transformation we come up with the Hamiltonian
\begin{eqnarray}
    H &=& \frac{1}{2} \mathbf{q}^T \mathbf{C}^{-1} \mathbf{q} + V \nonumber \\
    &=& 4\sum_{\substack{i, j= \\ 1,+,c,2}} E_{C,ij}n_i n_j + V \ ,
\end{eqnarray}
where $\mathbf{E_C} = e^2/2\cdot \mathbf{C}^{-1}$ denotes the charging energies and $n_i = q_i/2e$ are the charge operators normalized by the charge of a Cooper pair. In a system, where $E_C/E_J \ll 1$ we can treat the Hamiltonian as a sum of duffing oscillators, which can be quantized using 
\begin{eqnarray}
    n_i = in_i^{\text{ZPF}} \left( b_i - b_i^\dagger \right) \\
    \phi_i = \phi_i^{\text{ZPF}} \left( b_i + b_i^\dagger \right)
\end{eqnarray}
perturbed by small interactions. The $+$-mode is fully decoupled from the system due to the absence of a restricting potential and can be left out. The Hamiltonian reads
\begin{eqnarray}
    H &=& \sum_{i = 1, c, 2} \omega_i b_i^\dagger b_i + \frac{U_i}{2} b_i^\dagger b_i^\dagger b_i b_i + K_c \left( b_c^\dagger b_c^\dagger b_c + b_c^\dagger b_c b_c \right)\nonumber \\
    &-& \sum_{i = 1, 2} g_{ic} \bigl( b_i - b_i^\dagger \bigr) \bigl( b_c - b_c^\dagger \bigr) \nonumber \\
    &-& g_{12} \bigl( b_1 - b_1^\dagger \bigr) \bigl( b_2 - b_2^\dagger \bigr) \ ,
\end{eqnarray}
where the constant $K_c$ comes due to the non-vanishing third order contribution in the Taylor expansion of the coupler potential. The qubit level spacing is defined by
\begin{eqnarray}
    \omega_i &=& \sqrt{8E_{C,ii}E_{J_i}^{\text{eff}}} + U_i \\
    U_i &=& - E_{C,ii} \ .
\end{eqnarray}
and the coupling constants are given by 
\begin{equation}
    g_{ij} = 8E_{C,ij} \, n_i^{\text{ZPF}}n_j^{\text{ZPF}} \ .
\end{equation}

\section{\label{sec:SWtrafo} Schrieffer-Wolff transformation}
The goal of the Schrieffer-Wolff transformation is to transform the Hamiltonian in such a way that the coupling terms between the qubits and the coupler vanish in leading order. For this purpose we divide the Hamiltonian into 
\begin{eqnarray}
    H_0 &=& \sum_{i=1,c,2} \omega_i b_i^\dagger b_i + \frac{U_i}{2} b_i^\dagger b_i^\dagger b_i b_i - g_{12} \bigl( b_1 - b_1^\dagger \bigr)\bigl( b_2 - b_2^\dagger \bigr) \nonumber \\
    H_1 &=& - \sum_{i=1, 2} g_{ic} \bigl( b_i - b_i^\dagger \bigr)\bigl( b_c - b_c^\dagger \bigr)
\end{eqnarray}
and transform it via $H' = e^SHe^{-S}$, where the antihermitian operator $S$ needs to fulfill the condition 
\begin{equation}
H_1 + \left[ H_0, S \right] = 0 \ ,
\end{equation}
which is satisfied by the ansatz 
\begin{equation}
\label{eqn:Sansatz}
    S = \sum_{i=1, 2} \frac{g_{ic}}{\Delta_{ic}} \bigl( b_i^\dagger b_c - b_c^\dagger b_i \bigr) - \frac{g_{ic}}{\Sigma_{ic}} \bigl( b_i^\dagger b_c^\dagger - b_c b_i \bigr) \ ,
\end{equation}
when neglecting higher order terms as well as terms that do not affect the dynamics in the lowest lying two-level subsystem. Here $\Sigma_{ic} = \omega_i + \omega_c$ and $S$ is chosen in such a way that the coupling processes between the qubits and the coupler vanish in lowest order. 
The effective Hamiltonian is then given by
\begin{equation}
    H_{\text{eff}} = H_0 + \frac{1}{2} \left[ H_1, S\right] \ .
\end{equation}
One can simplify the last equation by using that the coupler always stays approximately in its groundstate. For $g_{ic} \ll \Delta_{ic}$ and $g_{ic} \ll \Sigma_{ic}$ one can expand the transformed Hamiltonian in orders of $g_{ic} / \Delta_{ic}$ and $g_{ic} / \Sigma_{ic}$. To leading order this results in the effective two qubit Hamiltonian
\begin{equation}
    H_{\text{eff}} = \sum_{i=1, 2} \tilde{\omega}_i b_i^\dagger b_i + \frac{U_i}{2} b_i^\dagger b_i^\dagger b_i b_i - g_{\text{eff}} \bigl( b_1 - b_1^\dagger \bigr) \bigl( b_2 - b_2^\dagger \bigr) 
\end{equation}
with the effective qubit frequencies $\tilde{\omega}_i = \omega_i + g_{ic}^2(1/\Delta_{ic} - 1/\Sigma_{ic})$ and the effective coupling constant 
\begin{equation}
\label{eqn:SW}
    g_{\text{eff}} = g_{12} + \frac{g_{1c}g_{2c}}{2} \left( \frac{1}{\Delta_{1c}} + \frac{1}{\Delta_{2c}} - \frac{1}{\Sigma_{1c}} - \frac{1}{\Sigma_{2c}} \right) \ ,
\end{equation}
which can be fully suppressed by tuning the coupler frequency $\omega_c$ appropriately. Since nevertheless $|\Sigma_{1c}|, |\Sigma_{2c}| \gg |\Delta_{1c}|, |\Delta_{2c}|$, the coupling is fully suppressed if condition (\ref{eqn:SW_cond}) holds.

\section{\label{sec:statelocalization} Localization of the qubit states}
In this section we will show, that the qubit delocalization parameter $\epsilon$
\begin{equation}
    \epsilon = \text{max} \left( \left| \langle \widetilde{100} | 001 \rangle \right|^2 \ , \ \left| \langle \widetilde{001} | 100 \rangle \right|^2 \right)
\end{equation}
(cf. \autoref{eqn:epsilon}) is minimized for the vanishing coupling condition $g_{\text{eff}} = 0$ calculated from a Schrieffer-Wolff transformation. 

First, we calculate the state corrections of the qubit states using perturbation theory, considering the coupling terms as a small perturbation to the qubit level spacings. The dressed states at leading order are given by
\begin{eqnarray}
    \label{eq:pert_comp_basis}
    \ket{\widetilde{100}} &=& \ket{100} + \frac{g_{1c}}{\Delta_{1c}} \ket{010} \nonumber\\
    &+& \frac{1}{\Delta_{12}}\left( g_{12} + \frac{g_{1c}g_{2c}}{\Delta_{1c}}\right) \ket{001}\\
    \ket{\widetilde{001}} &=& \ket{001} + \frac{g_{2c}}{\Delta_{2c}} \ket{010} \nonumber\\
    &+& \frac{1}{\Delta_{21}}\left( g_{12} + \frac{g_{1c}g_{2c}}{\Delta_{2c}}\right) \ket{100} \ .
\end{eqnarray} 
Thus, the qubit delocalization can be approximated by
\begin{eqnarray}
    \epsilon &=& \frac{1}{\Delta_{12}^2} \ \text{max} \left( \left( g_{12} + \frac{g_{1c}g_{2c}}{\Delta_{1c}}\right)^2 \ , \ \left( g_{12} + \frac{g_{1c}g_{2c}}{\Delta_{2c}}\right)^2 \right) \nonumber \\
    &:=& \text{max} \left( f_1 \ , \ f_2\right) \ ,
\end{eqnarray}
where $f_i(\omega_c) = 1/\Delta_{12}^2 \cdot (g_{12} + g_{1c}g_{2c}/\Delta_{ic})^2$. Since $\epsilon$ is positive and continuous away from the poles, i.e. where perturbation theory is valid, the intersections of $f_i(\omega_c)$ can be associated with local minima
\begin{eqnarray}
    0 &\overset{!}{=}& f_1(\omega_c) - f_2(\omega_c) \notag \\
    \Leftrightarrow 0 &=& 2g_{12}g_{1c}g_{2c} \left( \frac{1}{\Delta_{1c}} - \frac{1}{\Delta_{2c}}\right) + g_{1c}^2g_{2c}^2 \left( \frac{1}{\Delta_{1c}^2} - \frac{1}{\Delta_{2c}^2}\right) \notag \\
    \Leftrightarrow 0 &=& g_{12} + \frac{g_{1c}g_{2c}}{2}\left( \frac{1}{\Delta_{1c}} + \frac{1}{\Delta_{2c}} \right) = g_{\text{eff}} \ .
\end{eqnarray}
In order to identify the global minimum of $\epsilon$, we first calculate the zeros of $g_{\text{eff}}(\omega_c)$ and then put the values we find into $f_1(\omega_c)$ or $f_2(\omega_c)$. The zeros of $g_{\text{eff}}$ are given by
\begin{equation}
    \omega_c^\pm = \frac{1}{2} \left( \Sigma_{12} + \frac{g_{1c}g_{2c}}{g_{12}} \pm \sqrt{\Delta_{12}^2 + \frac{g_{1c}^2g_{2c}^2}{g_{12}^2}}\right) \ .
\end{equation}
We can now calculate $\epsilon(\omega_c^\pm)$ to find the best separation of the qubit eigenstates:
\begin{equation}
    \epsilon(\omega_c^\pm) = \left( \frac{g_{12}}{\frac{g_{1c}g_{2c}}{g_{12}} \pm \sqrt{\Delta_{12}^2+\frac{g_{1c}^2g_{2c}^2}{g_{12}^2}}}\right)^2
\end{equation}
We set without loss of generality $\Delta_{12}>0$ and distinguish the cases $\text{sgn}(g_{12}g_{1c}g_{2c}) = 1$ and $\text{sgn}(g_{12}g_{1c}g_{2c}) = -1$. For the former $\epsilon(\omega_c^+)<\epsilon(\omega_c^-)$, and thus the best qubit localization is reached for $\omega_c$ above the two qubit frequencies:
\begin{eqnarray}
    \omega_c &=& \frac{1}{2} \left( \Sigma_{12} + \frac{g_{1c}g_{2c}}{g_{12}} + \sqrt{\Delta_{12}^2+\frac{g_{1c}^2g_{2c}^2}{g_{12}^2}} \right) \nonumber \\
    &>& \frac{1}{2} \left( \Sigma_{12} + \Delta_{12} \right) = \omega_1 \ .
\end{eqnarray}
For the latter case, we find the best qubit localization for $\epsilon(\omega_c^-)<\epsilon(\omega_c^+)$, so that $\omega_c$ is below the qubit frequencies:
\begin{eqnarray}
    \omega_c &=& \frac{1}{2} \left( \Sigma_{12} + \frac{g_{1c}g_{2c}}{g_{12}} - \sqrt{\Delta_{12}^2+\frac{g_{1c}^2g_{2c}^2}{g_{12}^2}} \right) \nonumber \\ 
    &<& \frac{1}{2} \left( \Sigma_{12} - \Delta_{12} \right) = \omega_2 \ .
\end{eqnarray}

\section{\label{sec:ZZcouplingPert} Perturbative calculation of the $ZZ$-coupling}
The static $ZZ$-coupling can be calculated using perturbation theory as described in \autoref{sec:ZZcoupling}. The relevant terms are given by 
\begin{eqnarray}
    \zeta^{(0)} =& & \zeta^{(1)} = 0 \notag \\
    \zeta^{(2)} =& & g_{12}^2 \left( \frac{2}{\Delta_{12} - U_2} + \frac{2}{\Delta_{21} - U_1}\right) \notag \\
    \zeta^{(3)} =& &g_{12}g_{1c}g_{2c} \biggl( \frac{4}{(\Delta_{21} - U_1)\Delta_{2c}} + \frac{4}{(\Delta_{12} - U_2)\Delta_{1c}} \notag \\
    & &+ \frac{2}{\Delta_{1c}\Delta_{2c}} - \frac{2}{\Delta_{21}\Delta_{2c}} - \frac{2}{\Delta_{12}\Delta_{1c}}\biggr) \notag\\
    \zeta^{(4)} =& & g_{1c}^2g_{2c}^2 \Biggl( \left( \frac{1}{\Delta_{1c}} + \frac{1}{\Delta_{2c}} \right)^2 \frac{2}{\Delta_{1c}+\Delta_{2c}-U_c} \notag \\
    & &- \biggl(\frac{1}{\Delta_{1c}}\biggr)^2 \left( \frac{1}{\Delta_{12}} + \frac{1}{\Delta_{2c}} - \frac{2}{\Delta_{12} - U_2}\right) \notag \\
    & &- \biggl(\frac{1}{\Delta_{2c}}\biggr)^2 \left( \frac{1}{\Delta_{21}} + \frac{1}{\Delta_{1c}} - \frac{2}{\Delta_{21} - U_1}\right)\Biggr) \notag
\end{eqnarray}
The total $ZZ$-coupling to leading order is then given by the sum of these terms
\begin{equation}
    \zeta = \sum_i \zeta^{(i)} \ .
\end{equation}

\section{\label{sec:ZZsuppression1} Suppression of residual couplings with an anharmonicity asymmetry in the qubits}
Following the procedure in \autoref{sec:cross_analysis} we can derive the sweet spot for the coupler characteristics also for Transmons of different anharmonicities. We introduce therefore the anharmonicity asymmetry parameter $\delta$ and define the qubit anharmonicities to be
\begin{eqnarray}
    U_1 &=& (1+\delta) U \nonumber \\
    U_2 &=& (1-\delta) U \ .
\end{eqnarray}
Setting \autoref{eqn:zeta_perturbation} to zero and using condition \ref{eqn:SW_cond} we come up with an extended equation for the perfect coupler anharmonicity at the idling point:
\begin{widetext}
\begin{equation}
\label{eqn:Uc_extended}
    U_c = -\frac{U}{2} \left( 1 - \frac{U^2}{\Delta_{12}^2} - \frac{U}{2(\Delta_{1c} + \Delta_{2c})} + \frac{2\delta U}{\Delta_{12}} + \frac{2\delta^2U^2}{\Delta_{12}^2}\right)^{-1} \ .
\end{equation}
\end{widetext}
In the limit $\delta \rightarrow 0$ the formula reduces to the one given in \autoref{eqn:Uc_lowcrosstalk}. Note that the anharmonicity of the C-Shunt flux coupler in our simulations differs by a few MHz compared to the one which would be expected by \autoref{eqn:Uc_extended}. This is due to counter rotating wave terms in the Hamiltonian, which are neglected for simplicity in the theoretical derivation of this formula.

\section{\label{sec:parameters_detailed} Full simulation parameters}
The parameters for the simulations shown in this paper are oriented to state of the art quantum hardware, but are freely chosen. The capacities are listed in \autoref{tab:capacitances}.
\begin{table}
    \centering
    \begin{tabular}{p{1cm} p{1cm} p{1cm} p{1cm} p{1cm} p{1cm} p{1cm}}
         \toprule 
         \begin{center} \vspace{-0.5cm} $C_1$ \end{center} & \begin{center} \vspace{-0.5cm} $C_2$ \end{center} & \begin{center} \vspace{-0.5cm} $C_C$ \end{center} & \begin{center} \vspace{-0.5cm} $C_{g}$ \end{center} & \begin{center} \vspace{-0.5cm} $C_{12}$ \end{center} & \begin{center} \vspace{-0.5cm} $C_{1c}$ \end{center} & \begin{center} \vspace{-0.5cm} $C_{2c}$ \end{center} \\[-1.7ex] \hline 
         \begin{center} \vspace{-0.4cm} 85 \end{center} & \begin{center} \vspace{-0.4cm} 85 \end{center} & \begin{center} \vspace{-0.4cm} 30 \end{center} & \begin{center} \vspace{-0.4cm} 70 \end{center} & \begin{center} \vspace{-0.4cm} 0.23 \end{center} & \begin{center} \vspace{-0.4cm} 7.9 \end{center} & \begin{center} \vspace{-0.4cm} 7.9 \end{center} \\[-2.3ex] \bottomrule
    \end{tabular}
    \caption{Capacitances used for the simulation of the system. All values are given in $\mathrm{fF}$.}
    \label{tab:capacitances}
\end{table}
The qubits were set at the idling point to 
\begin{figure}[!htp]
    \centering
    \includegraphics[width = \columnwidth]{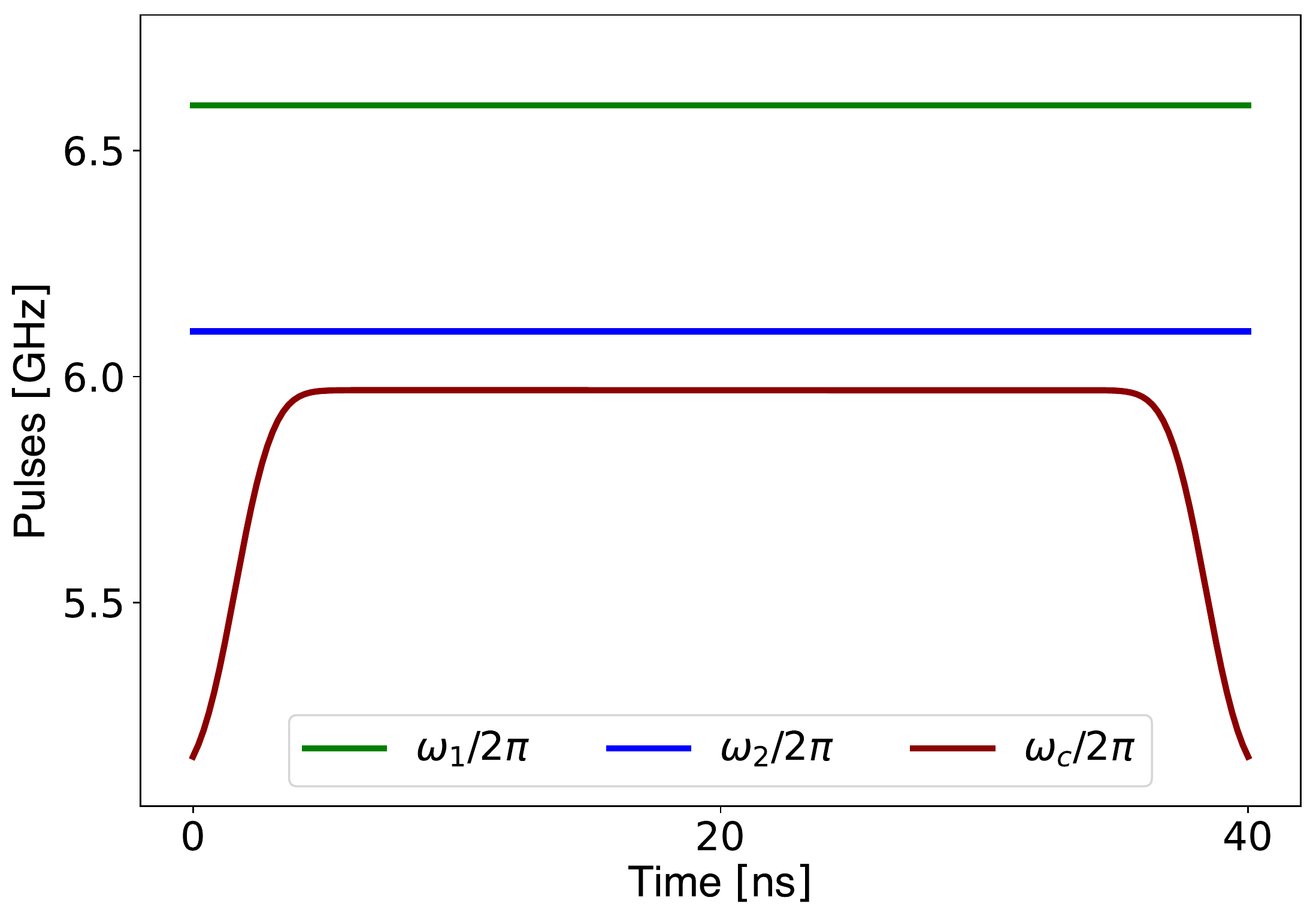}
    \caption{Pulse shapes for the tuning of coupler frequency. Tuning the coupler frequency activates a $ZZ$-interaction between the qubits during the gate.}
    \label{fig:pulses}
\end{figure}
\begin{eqnarray}
    \omega_1/2\pi &=& 6.6\,\mathrm{GHz} \nonumber \\
    \omega_2/2\pi &=& 6.1\,\mathrm{GHz} \\
    \omega_c/2\pi &=& 5.1\,\mathrm{GHz} \nonumber
\end{eqnarray}
The two Josephson junctions in the upper branch of the coupler have an energy of $E_J/h = 41.2\,\mathrm{GHz}$ measured in the $1/\sqrt{2}(\phi_{c1}-\phi_{c2})$-basis and the other Josephson junction in the lower branch is smaller by a factor $\alpha = 0.2347$. 

The couplings at the idling point are therefore given by 
\begin{eqnarray}
    g_{12}/2\pi &=& 14.32\,\mathrm{MHz} \nonumber \\
    g_{1c}/2\pi &=& 142.98\,\mathrm{MHz}  \\
    g_{2c}/2\pi &=& -137.63\,\mathrm{MHz} \nonumber
\end{eqnarray}
For the CZ-gate we use a gaussian flattop shaped pulse described by 
\begin{eqnarray}
    \omega_c (t) = \omega_{c, \, \text{idle}} &+& \frac{\omega_{c, \, \text{int}} - \omega_{c, \, \text{idle}}}{4} \bigl( 1 + \text{erf} \left( (t - \tau) / \tau \right) \bigr) \cdot \nonumber \\
    & &\bigl( 1 + \text{erf} \left( (t_{\text{gate}} - t - \tau) / \tau \right) \bigr)
\end{eqnarray}
to tune the system away from its idling point $\omega_{\text{idle}}$ to an interaction point $\omega_{\text{int}}$. We optimized the risefall parameter $\tau$ as well as the amplitude of the envelope $\omega_{\text{int}} - \omega_{\text{idle}}$ for the coupler. The pulses for the optimized paramters are depicted in \autoref{fig:pulses}.

\section{\label{sec:CZ_fast_ad} Fast adiabatic gate scheme}
\begin{figure}[!htp]
    \centering
    \includegraphics[width = \columnwidth]{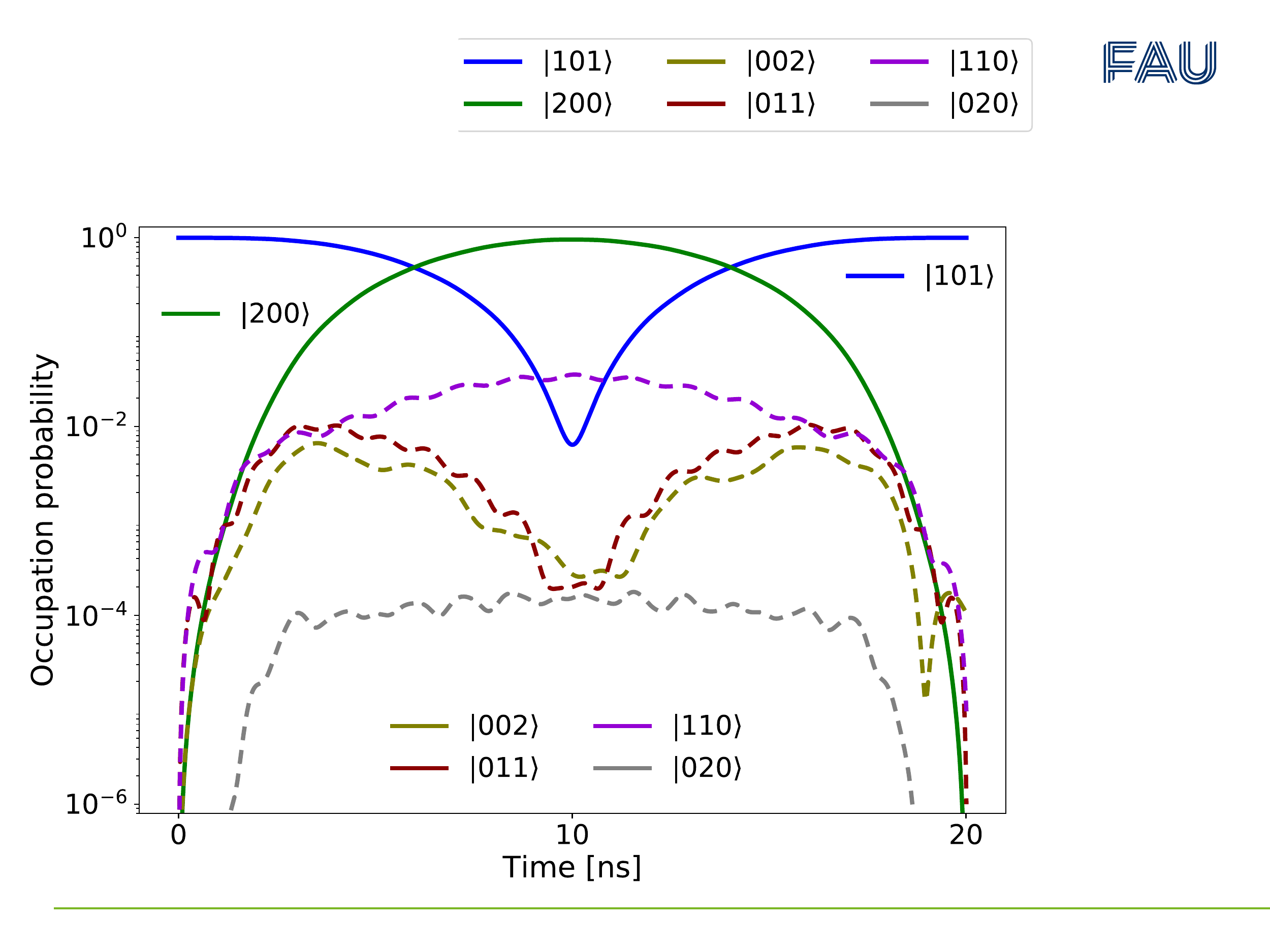}
    \caption{Simulation of a CZ-gate on the presented tunable coupler scheme. The curves show the population probability of the energetically adjacent states of $\ket{\widetilde{101}}$ during the gate. Population of these states at $t=t_{\text{gate}}$ is responsible for leakage errors.}
    \label{fig:CZ_gate}
\end{figure}

Another way to implement the CZ-gate is to bring the states $\ket{\widetilde{101}}$ and $\ket{\widetilde{200}}$ into resonance.
If one now activates an effective coupling between the two states by increasing the coupler frequency, they start to oscillate. If one waits for a complete oscillation, the state $\ket{\widetilde{101}}$ collects the phase $\pi$ which is characteristic for the CZ-gate. A plot of the state populations for such a gate scheme can be seen in \autoref{fig:CZ_gate}. 

\begin{figure}
    \centering
    \hspace{0cm}
    \includegraphics[width = \columnwidth]{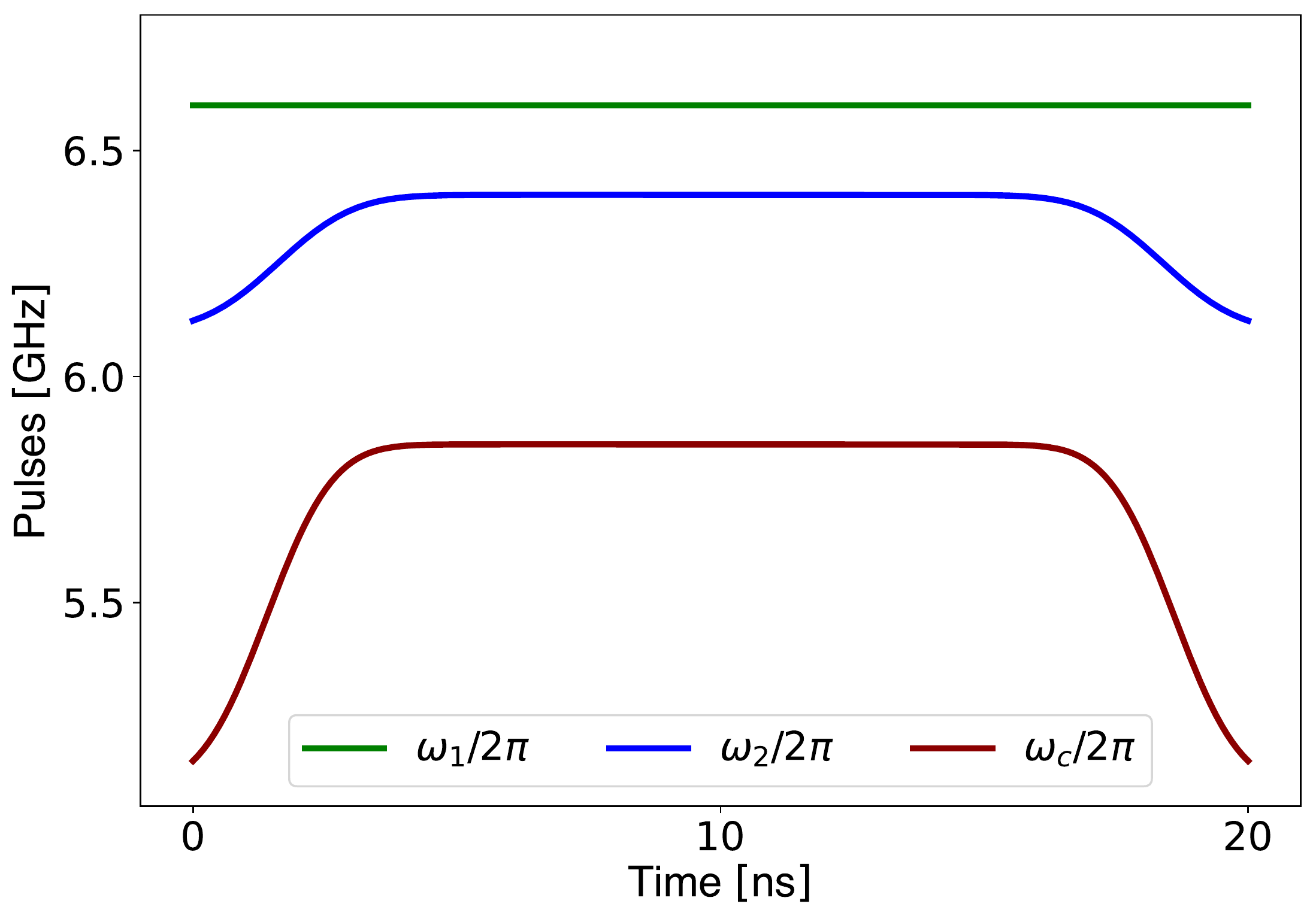}
    \caption{Pulse shapes for the tuning of the qubit and coupler frequencies.}
    \label{fig:pulses2}
\end{figure}

For this gate we also use gaussian flattop curves which are shown in \autoref{fig:pulses2}. In total we simulate a gate with a length of $20\,\mathrm{ns}$. The process infidelity can be approximated to $\varepsilon \approx 3\cdot 10^{-5}$, which is an order of magnitude smaller than the error due to decoherence.

\FloatBarrier

\nocite{*}
\bibliography{mybib}

\end{document}